\documentclass[10pt,prl,superscriptaddress,showpacs]{revtex4}
\usepackage{amsmath}
\usepackage{amsfonts}
\usepackage{amssymb}
\usepackage{graphicx}
\newcommand{\ket}[1]{\left|\, #1 \right\rangle}

\newcommand{\bra}[1]{\left \langle \, #1 \right|}

\newcommand{\mean}[1]{\left\langle \,  #1 \, \right\rangle}
\begin{document}
\title{Analysis of critical parameters in the scheme of Bj\"ork, Jonsson, and S\'anchez-Soto}
\author{Marcin Wie\'sniak}\affiliation{Instytut Fizyki Teoretycznej i Astrofizyki,\\ Uniwersytet Gda\'nski, PL-80 952 Gda\'nsk, Poland}\affiliation{Department of Physics, National University of Singapore, Singapore 117542}
\author{Marek \.Zukowski}\affiliation{Instytut Fizyki Teoretycznej i Astrofizyki,\\ Uniwersytet Gda\'nski, PL-80 952 Gda\'nsk, Poland}

\begin{abstract}
Bj\"ork, Jonsson, and S\'anchez-Soto describe an interesting
(gedanken-)experiment which
demonstrates that single photons can indeed lead to
effects which have no local realistic description. 
We study the
critical values of parameters of some possible features of a non-perfect realisation of the experiment (especially photon loss, which could be looked at as the
detection efficiency), that need to be satisfied so that the
experiment can be considered as a valid test of quantum mechanics
versus local realism. Interestingly, the scheme turns out to be 
robust against photon loss.
\end{abstract}
\maketitle

Not only is the Bell theorem \cite{Bell} related to foundations of
physics, but also to advanced (quantum) information processing
tasks. It allows to exclude all theories based on
local hidden variables experimentally. Up to date, there have been many realizations of a Bell-type experiments \cite{aspect,weihs,rowe}, none of which 
did close all the possible loopholes. The most conspirative theory would allow nature to choose in which loophole local realism can hide from the observers' perception. Therefore, ever since the pioneer
attempts of falsification of local realism, the results always left
some doubts. In early experiments (see e.g. \cite{aspect}) the emitted
light was not correlated directionally, because a calcium atom
cascade was used as a source. It emits the photons in random
directions. In the scheme of Weihs {\em et al.} \cite{weihs}, which
was a parametric down-conversion refinement of the Aspect {\em et
al} experiment \cite{aspect}, it was for the first time possible to
close the locality loophole by changing the observables fast enough,
and locating the detection stations far enough from the source.
However, the main problem in optical realizations of EPR tests is
the detection efficiency. Experiments with entangled atoms allow for
much higher efficiency. However, in ref. \cite{rowe}, where
almost perfect detection efficiency was reported, the spatial
separation between the atoms was much to close to call the
experiment loophole free. The scheme of \cite{Bjork}, as we shall
see, lowers very much the efficiency requirements  in optical
Bell-type tests.

For the sake of the further consideration, we
begin with recalling how the transmission and detection efficiency enters the discussion on the falsification of local realism.
Clauser and Horne \cite{ch1} derived a Bell inequality for a
following experimental situation: two separated observers, say
Carol and Daniel, get particles from an entangled pair in a
singlet state $\ket{\Psi^-}\!=(\ket{01}-\ket{10}\!)/\sqrt{2}$. They
can, independently from each other, choose between two local states,
$(\ket{0}+e^{i\phi_k}\ket{1})/\sqrt{2}$ or
$(\ket{0}+e^{i\phi_k'}\ket{1})/\sqrt{2}$, ($k=c,d$) and observe 
detection events associated with one of these states. For phases $\phi_c$ and $\phi_d$ probabilities that they would succeed are denoted as
$P(\phi_c)$ and $P(\phi_d)$, respectively, and the joint probability
as $P(\phi_c,\phi_d)$. Were these probabilities described by any
local and realistic theory, the CH inequality
\begin{eqnarray}
\label{ch} P(\phi_c,\phi_d) +P(\phi_c,\phi'_d)
+P(\phi'_c,\phi_d)\nonumber\\-
P(\phi'_c,\phi'_d)\!-P(\phi_c)\!-P(\phi_d)&\!\leq 0
\end{eqnarray}
should hold.

We consider two kinds of imperfections of the setup, namely that the
detectors and transmission channels work with a finite efficiency $\eta$, and
depolarization, transforming the pure state
$\ket{\Psi^-}\bra{\Psi^-}$ into a mixture $l\!
\ket{\Psi^-}\bra{\Psi^-}\!+(1-l)\hat{I}_{2\times 2}/4 \!\quad(0\geq
l\geq 1)$, as in \cite{Werner}. Taking these two effects into
account we obtain that $P(\phi_k)\!=\eta/2(k=c,d)$,
$P(\phi_c,\phi_d)\!=\eta^2(1-l\cos(\phi_c-\phi_d))/4$, and similarly
for all other choices of phases. This implies a relation between the
critical efficiency and critical the depolarization parameter
$\eta_{CRIT}=2/(\sqrt{2}l_{CRIT}+1)$ (above the critical values of
{\em both} parameters the CH inequality can be violated).

Another possibility is to consider a Clauser-Horne-Shimony-Holt inequality \cite{chsh,garg}. Each observer (randomly) chooses one of two dichotomic observables ($C,C'$ for Charlie, $D,D'$ for Daniel) and measurement can yield one of two distinct results, $+1$ or $-1$. The correlation function is defined as a mean of a product of the two results over many runs of the experiment, $E(C,D)=\mean{CD}$. All local realistic theories imply that
\begin{equation}
\label{chsh1}
|E(C,D)+E(C,D')+E(C'D)-E(C'D')|\leq 2.
\end{equation}
Assuming the state to be $\rho=l\!
\ket{\Psi^-}\bra{\Psi^-}\!+(1-l)\hat{I}_{2\times 2}/4$ we get the correlation function as $E(X,Y)=-l\vec{x}\cdot\vec{y}$, where $X=\vec{x}\cdot\vec{\sigma}_c$ represents $C$ or $C'$ and, similarly, $Y=\vec{y}\cdot\vec{\sigma}_d$ stands for $D$ or $D'$. Here $\vec{\sigma}_k$ is a vector of Pauli matrices acting on the respective Hilbert space. For detectors with non-unit efficiency, we succeed to register a known result in only a fraction $\eta^2$ of all experimental runs. One can assign to the "no click" event the value $+1$, see \cite{garg}. The efficient correlation function is thus $E_{eff}(X,Y)=\eta^2E(X,Y)+(1-\eta)^2$. After putting it into (\ref{chsh1}) and some straightforward algebra, one gets the same critical relation between $l$ and $\eta$ as in case of the CH inequality.
Thus, in an experiment with two maximally entangled particles and two measurement settings a local
realistic description  cannot be convincingly
excluded without detectors with the efficiency below
$2/(\sqrt{2}+1)\approx 82.8\%$. 
Eberhard gave a proposal for a loophole free Bell experiment \cite{eberhard}, in which the required efficiency to violate CH inequalities can be as low as 66,7\%. This is done, however, with the help of non-maximally entangled states, and in fact in the limit of product statets. Can other possible realizations of a
Bell test allow to decrease this bound?

The scheme of \cite{Bjork} is a realization of
the ideas of Tan, Walls and Collett \cite{TAN}. One starts with  a
single photon with a $-45^\circ$
polarization, what we can write as:
\begin{eqnarray}
\frac{1}{\sqrt{2}}(\ket{H}-\ket{V})=\frac{1}{\sqrt{2}}(\hat{a}^\dagger_H-\hat{a}^\dagger_V)\ket{0,0}\nonumber\\
=\frac{1}{\sqrt{2}}(\ket{1,0}-\ket{0,1}).
\end{eqnarray}
The last equation is written using a version of the Fock space formalism in which the
photon is represented by a superposition of the first polarization mode (horizontal $H$)
in the single photon state and the second one (vertical $V$) in the vacuum
state, with  the $H$ mode in the vacuum state and $V$ in the single
photon state. 

The photon is sent to an input channel $a$ of the
PBS. A reference light from a local oscillator is added
through the second input channel $b$. The reference beam is
coherent, originally of a mean photon number 2$|\alpha|^2$
(hereafter, we take $\alpha$ real), and polarized at
$+45^{\circ}$. The PBS splits both signals into two channels $c$ and
$d$. During the propagation phase shifts $\omega\tau_c$ and $\omega\tau_d$ are picked ($\omega$ is the
frequency). At the end we have measuring devices. The setup is presented in figure 1:

\begin{figure}[h]
\centering
\includegraphics[width=7cm]{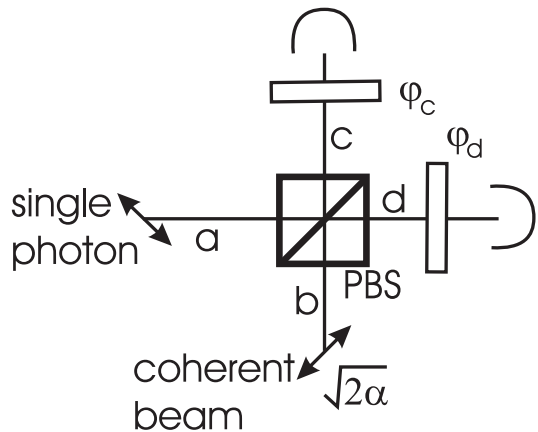}
\caption{The scheme of Bj\"ork, Jonsson, and S\'anchez-Soto. A single photon and the coherent beam are mixed on a polarizing beam-splitter (PBS). Each observer is seated at one output of PBS and makes specific measurements described in the main text. The measured observables depend on a local phase   $\phi_c$ and $\phi_d$. The measuring devices are just suggested (i.e., they are some black boxes which measure the required observables).} \label{xiang}
\end{figure}
Thus behind the PBS the state is
$\ket{\phi}\!=\!\frac{1}{\sqrt{2}}\!(e^{i\omega\tau_c}\ket{1,\alpha
e^{i\omega\tau_c},\alpha
e^{i\omega\tau_d},0}\!-\!e^{i\omega\tau_d}\ket{0,\alpha
e^{i\omega\tau_c},\alpha e^{i\omega\tau_d},1}),$ with mode ordering
$c_H,$ $c_V,$ $d_H,$ $d_V$ (for convenience and without a
loss of generality we choose $\omega\tau_c$ and $\omega\tau_d$ to be
multiples of $2\pi$), and the reduced state of modes of one of the outputs is
$\rho_k\!=(\ket{1}\bra{1}+\ket{0}\bra{0})\!\otimes\ket{\alpha}\bra{\alpha}/2$,
where the first Hilbert space refers to the single photon polarization and
the other--to the coherent state polarization. Measuring devices depend of a local macroscopic variable $\phi_k$, and  
should be able to detect $n_k$-photon states defined by
$\ket{+,n_k,\phi_k}=\left(1+\frac{n_k}{\alpha^2}\right)^{-1/2}\left(\sqrt{n_k}/\alpha\ket{0,n_k}+e^{i\phi_k}\ket{1,n_k-1}\right)$.
The probability of such an event is
$P_{+}(n_k,\phi_k)\!=e^{-\alpha^2}\alpha^{2(n_k-1)}\big/\left(\left(1+{n_k}/{\alpha^2}\right)(n_k-1)!\right)$.
The probabilities that would enter the inequalities are sums of probabilities of such events
\begin{eqnarray}
P_{+}(\phi_k)=\sum_{n_k=1}^\infty P_{+}(n_k,\phi_k),\\
P_{++}(\phi_c,\phi_d)=\sum_{n_k=1}^\infty \sum_{n_m=1}^\infty P_{++}(n_k,\phi_k,n_m, \phi_d).
\end{eqnarray}
In the ideal case one has
\begin{equation}
P_{++}(\phi_c,\phi_d)=2\sin^2(\phi_c-\phi_d)P_{+}(\phi_c)P_{+}(\phi_d).
\label{G}
\end{equation}
Since locally there is no dependence on the phase, using  the
relation (\ref{G}) one can show that 
Clauser-Horne inequality (\ref{ch})  can be violated whenever $P_{+}(\phi_k)>1/(1+\sqrt{2})$.

The authors of Ref. \cite{Bjork} stress that the observation of the
correlations is more efficient for a strong coherent field, with
$\alpha^2>>1$. Therefore we shall
discuss robustness of the setup against imperfections only for such fields.

An imperfect transmission \footnote{Since the workings of the measuring devices, which would be capable of performing the required tasks are not known, one cannot discuss the detection efficiency of such devices. Thus, we discuss only the transmission efficiency.}  with an efficiency $\eta$ is equivalent to a
perfect one with beam splitters, both of a transmittivity  $\eta$, put into outputs of PBS, but we neglect the signal reflected by them.
Its action on the coherent part of the state
preserves coherences but decreases the excitation number by a factor of $\eta$. The one-photon part is being
statistically mixed with vacuum, as we trace out external modes of
the field. The state becomes
\begin{eqnarray}
\frac{1}{2}(\ket{1,\alpha,\alpha,0}-\ket{0,\alpha,\alpha,1})(\bra{1,\alpha,\alpha,0}-\bra{0,\alpha,\alpha,1})&\rightarrow&\nonumber\\
\frac{\eta}{2}(\ket{1,\alpha\sqrt{\eta},\alpha\sqrt{\eta},0}-\ket{0,\alpha\sqrt{\eta},\alpha\sqrt{\eta},1})\nonumber\\
\times(\bra{1,\alpha\sqrt{\eta},\alpha\sqrt{\eta},0}-\bra{0,\alpha\sqrt{\eta},\alpha\sqrt{\eta},1})&+&\nonumber\\(1-\eta)\ket{0,\alpha\sqrt{\eta},\alpha\sqrt{\eta},0}\bra{0,\alpha\sqrt{\eta},\alpha\sqrt{\eta},0}.\nonumber\\
\label{XXX}
\end{eqnarray}
Note that what is important here is only the the attenuation of the single photon input. On can always increase the value of the initial amplitude of the coherent field to compensate the channel inefficiency. Nevertheless, we shall use the above approach of (\ref{XXX}).

 We can also introduce decoherence to our model. For
simplicity, we assume that only a (strongly non--classical)
single-photon part of the state is exposed to destructive interaction with the environment, while the
coherent part of the state remains unaffected. The loss of coherence can
be described by a transition:
\begin{eqnarray}
\frac{1}{2}\left(\ket{0,H_d}-\ket{V_c,0}\right)\left(\bra{0,H_d}-\bra{V_c,0}\right)&\rightarrow&\nonumber\\
l\frac{1}{2}\left(\ket{0,H_d}-\ket{V_c,0}\right)\!\left(\bra{0,H_d}-\bra{V_c,0}\right)
&+\nonumber\\
(1-l)\frac{1}{2}\left(\ket{0,H_d}\bra{0,H_d}+\ket{V_c,0}\bra{V_c,0}\right),
\end{eqnarray}
with the decoherence parameter $0\geq l\geq 1$. Then the global and
the reduced states become:
\begin{eqnarray}
\rho(\eta,l)=&\frac{l\eta}{2}(\ket{0,\alpha\sqrt{\eta},\alpha\sqrt{\eta},1}-\!\ket{1,\alpha\sqrt{\eta},\alpha\sqrt{\eta},0})\nonumber\\&\underbrace{\times(\bra{0,\alpha\sqrt{\eta},\alpha\sqrt{\eta},1}-\bra{1,\alpha\sqrt{\eta},\alpha\sqrt{\eta},0})}_{\quad single\quad photon\quad not\quad lost\quad and\quad coherent}\nonumber\\
+&\frac{(1-l)\eta}{2}(\ket{0,\alpha\sqrt{\eta}\,\alpha\sqrt{\eta},1}\bra{0,\alpha\sqrt{\eta}\,\alpha\sqrt{\eta},1}\nonumber\\&\underbrace{+\ket{1,\alpha\sqrt{\eta}\,\alpha\sqrt{\eta},0}\bra{1,\alpha\sqrt{\eta}\,\alpha\sqrt{\eta},0})}_{
single\quad photon\quad not\quad lost, \quad not \quad
coherent}\nonumber\\+&\underbrace{(1-\eta)\!\ket{0,\alpha\sqrt{\eta},\alpha\sqrt{\eta},0}\bra{0,\alpha\sqrt{\eta},\alpha\sqrt{\eta},0}}_{single\quad
photon\quad lost}
\end{eqnarray}
and
\begin{eqnarray}
\rho_{c(d)}=&\left(\frac{\eta}{2}\ket{1}\bra{1}+\left(1-\frac{\eta}{2}\right)\ket{0}\bra{0})\right)\nonumber\\
&\otimes\ket{\alpha\sqrt{\eta}}\bra{\alpha\sqrt{\eta}},
\end{eqnarray}
what results in the following probabilities:
\begin{eqnarray}
P_{+}(n_k,\phi_k)={e^{-\alpha^2\eta}\eta(3-\eta)(\alpha^2\eta)^{n_k-1}}\nonumber\\ \times\left(2\left(1+\frac{n_k}{\alpha^2}\right)(n_k-1)!\right)^{-1},\\
P_{++}(n_c,\phi_c,n_d,\phi_d)\nonumber\\=\frac{e^{-2\alpha^2\eta}(\alpha^2\eta)^{n_c+n_d-2}}{\left(1+\frac{n_c}{\alpha^2}\right)\left(1+\frac{n_d}{\alpha^2}\right)(n_c-1)!(n_d-1)!}\nonumber\\
\times\left(\frac{\eta l\left(
1+\eta\right)^2}{2}\sin^2\frac{\phi_c-\phi_d}{2}+(1-l)\eta+\left(1-\eta\right)\eta^2\right).\nonumber\\
\end{eqnarray}

The probabilities, that we have to sum up over $n_k$, are products of a function of $n_k$ and an element of the Poisson distribution, with $\alpha^2$ as the mean value. The distribution has the property that the variance $\mean{(n_k-\mean{n_k})^2}$ is equal to the mean value, $\mean{n_k}$. Taking $\alpha^2$ much larger than 1, one gets $\mean{n_k}$ neglible against $\mean{n_k}^2$ and $\mean{n_k^2}$, and hence the latter two may be taken equal. One can also draw similar arguments for higher moments being close to powers of the mean. For large $\alpha$ we thus take $\langle f(n_k)\rangle=f(\langle n_k\rangle)$ for any sufficiently smooth function $f$. In particular, we will use the following approximations:
\begin{equation}
\label{approx1} 
\sum_{n=1}^{\infty}\frac{1}{1+\frac{n}{\alpha^2}}e^{-\alpha^2x}\frac{(\alpha^2x)^{n-1}}{(n-1)!}\approx\frac{1}{1+x},
\end{equation}
\begin{equation}
\label{approx2}
 \sum_{n=1}^{\infty}\frac{1}{1+\frac{n}{\alpha^2}}e^{-\alpha^2x}\frac{n(\alpha^2x)^n}{n!\alpha^2}\approx\frac{x}{1+x},
\end{equation}
\begin{equation}
\label{approx3} 
\sum_{n=1}^{\infty}\frac{1}{1+\frac{n}{\alpha^2}}e^{-\alpha^2x}\frac{n(\alpha^2x)^{n-1}}{(n-1)!\alpha^2}\approx\frac{x}{1+x},  
\end{equation}
\begin{equation}
\label{approx4}
\sum_{n=1}^{\infty}\frac{1}{1+\frac{n}{\alpha^2}}e^{-\alpha^2x}\frac{(\alpha^2x)^n}{n!}\approx\frac{1}{1+x},
\end{equation}
with $0\leq x\leq 1$. Strictly speaking, in (\ref{approx1}-\ref{approx4}) we demand $\alpha^2x$, rather than $\alpha^2$ itself to be large. 
In figure 2 we compare the numerical values of the sums in ratios to their estimated values computed for $x=0.2$ Higher values of $x$ would increase the accuracy of the approximations. 
\begin{figure}[h]
\centering
\includegraphics[width=7cm]{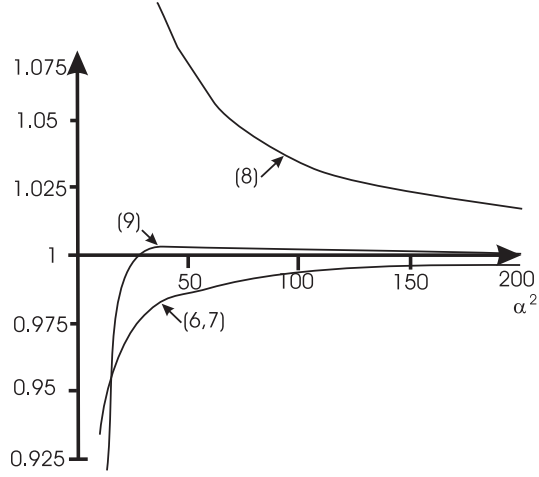}
\caption{Ratios between numerical values of left-hand sides of (\ref{approx1}-\ref{approx4}) and their estimated values as functions of $\alpha^2$ for $x=0.2$.}
\end{figure}

Using (\ref{approx1},\ref{approx2}) we get
$P_{+}(\phi_k)\approx\eta(3-\eta)/(2(1+\eta))$ and
$P_{++}(\phi_c,\phi_d)\!\approx\!
(2-\eta-l\cos(\phi_c-\phi_d))(\eta/(1+\eta))^2$
We can now put these probabilities into the CH inequality (\ref{ch})
and perform obvious steps. The first one is to choose the
optimal phases for the observers, such that $-\cos(\phi_c-\phi_d)-\cos(\phi_c-\phi'_d)-\cos(\phi'_c-\phi_d)+\cos(\phi_c'-\phi_d')=2\sqrt{2}$.
Next, to find
the critical values of $l$ and $\eta$, we set the Clauser-Horne
expression equal to zero and get
\begin{eqnarray}
\frac{-\eta_{CRIT}^3+2\eta_{CRIT}^2l_{CRIT}(1+\sqrt{2})-3\eta_{CRIT}}{(1+\eta_{CRIT})^2}=0,
\end{eqnarray}
which can be simplified to $l_{CRIT}=(3-2\eta_{CRIT}+\eta_{CRIT}^2)/(2\sqrt{2}\eta^2_{CRIT})$.

If the single photon can reach the measuring devices without a loss
of coherence $(l_{CRIT}=1)$, the critical transmission efficiency is
$\eta_{CRIT}\!=1+\sqrt{2}-2^{3/4}\!\approx
73.2\%$, while for perfect detectors the decoherence parameter should
be higher than $\frac{1}{\sqrt{2}}$. This indicates a great
similarity between decoherence of a single-photon state and
depolarization acting on a two-qubit state \cite{Werner}. Complete
decoherence of the single photon maps a state
$\frac{1}{2}(\ket{0,H_d}-\ket{V_c,0})(\bra{0,H_d}-\bra{V_c,0})$ onto
a ``classically correlated" (in the Fock space) mixture
$\frac{1}{2}(\ket{0,H_d}\bra{0,H_d}+\ket{V_c,0}\bra{V_c,0})$ rather
than the maximally mixed state, but since we make measurements in
bases, which are unbiased to the eigenbasis of this mixture, these
"classical correlations" play no role in the statistics. 

One can also consider the violation of the CHSH inequality \cite{chsh} when the described imperfections are taken into account. To construct the the correlation function we associate the states $\ket{+_{n_k,k}}=\frac{1}{\sqrt{1+\frac{n_k}{\alpha^2}}}\left(\frac{\sqrt{n_k}}{\alpha}\ket{0,n_k}+e^{i\phi_k}\ket{1,n_k-1}\right)$ with local outcomes $+1$ and $\ket{-_{n_k,k}}=\frac{1}{\sqrt{1+\frac{n_k}{\alpha^2}}}\left(\ket{0,n_k}-e^{i\phi_k}\frac{\sqrt{n_k}}{\alpha}\ket{1,n_k-1}\right)$ with $-1$. Its easy to show that the the sates span indeed the whole Hilbert space, except for the vacuum field. The projections $\ket{+,n_k,\phi_k}\bra{+,n_k,\phi_k}+\ket{+,n_k,\phi_k}\bra{+,n_k,\phi_k}$ is the identity operator acting on the subspace of local $n_k$-photon states. Obviously, summed over $n_k$ the projections constitute the global identity operator, except for the subspace of the vacuum.  The correlation function naively obtained from respective probabilities $E(\eta,l,\phi_c,\phi_d)=P_{++}(\eta,l,\phi_c,\phi_d)-P_{-+}(\eta,l,\phi_c,\phi_d)-P_{+-}(\eta,l,\phi_c,
\phi_d)+P_{--}(\eta,l,\phi_c,\phi_d)$, reads $E(\eta,l,\phi_c,\phi_d)=((1-\eta)/(1+\eta))^2(1-2\eta)+((2\eta)/(1+\eta))^2\cos(\phi_c-\phi_d)$. The CHSH inequality,
\begin{equation}
\label{chsh}
|E(\eta,l,\phi_c,\phi_d)+E(\eta,l,\phi_c,\phi_d')+E(\eta,l,\phi_c',\phi_d)-E(\eta,l,\phi_c',\phi_d')|\leq 2,
\end{equation} 
can be violated if
\begin{equation}
l>\frac{-\eta^3+3\eta^2-\eta+1}{2\sqrt{2}\eta^2}.
\label{forchsh}
\end{equation}
If the system preserves the perfect coherence, the critical efficiency is found to be $\eta_{CRIT}'=(3\sqrt{2})/(4+\sqrt{2})\approx 71.8\%$. As before, the inequality can be violated only if $l>1/\sqrt{2}$.

These two results cannot be mutually consistent. The CHSH inequality can be expressed as a combination of CH expressions and thus it is less general. On the other hand, we have obtained that the CH inequality require finer experimental conditions than CHSH. Thus a closer analysis of the problem  must allow the CH inequality to be violated even with less efficient channels. 

In order to achieve this, both Charlie and Daniel must have more freedom than just changing relative phases $\phi_c$ $\phi_d$ in (\ref{ch}). Let us allow them the following. If they set their local phase to the unprimend value, they should monitor successful local projections onto $\sum_{n_k=1}^\infty\ket{+,n_k,\phi_k}\bra{+,n_k,\phi_k}$, whereas once they choose the primed phases the count events are related to successful projections onto $\sum_{n_k=1}^\infty\ket{-,n_k,\phi'_k}\bra{-,n_k,\phi'_k}$. The new probalilities read

\begin{eqnarray}
P_{++}(\phi_c,\phi_d)=\left(\frac{\eta}{1+\eta}\right)^2(2-\eta-l\cos(\phi_c-\phi_d)),\nonumber\\
P_{-+}(\phi'_c,\phi_d)=\left(\frac{\eta}{(1+\eta)^2}\right)(\eta^2-\eta+2+l\cos(\phi'_c-\phi_d)),\nonumber\\
P_{+-}(\phi_c,\phi'_d)=\left(\frac{\eta}{(1+\eta)^2}\right)(\eta^2-\eta+2+l\cos(\phi_c-\phi'_d)),\nonumber\\
P_{--}(\phi_c,\phi_d)=\left(\frac{\eta}{1+\eta}\right)^2(1-l\cos(\phi_c-\phi_d)).
\end{eqnarray}
These probabilities, put into (\ref{ch}):
\begin{eqnarray}
\label{ch2}
P_{++}(\phi_c,\phi_d)+P_{+-}(\phi_c,\phi'_d)+P_{-+}(\phi'_c,\phi_d)&\nonumber\\
-P_{--}(\phi'_c,\phi'_d)-P_+(\phi_c)-P_+(\phi_d)\leq 0,
\end{eqnarray}
 yield that local realistic theories can be excluded only if $l>\frac{3-\eta}{2\sqrt{2}}$. In the extreme case of $l=1$, the Bell inequality can be thus violated for $\eta>3-2\sqrt{2}\approx 17.15\%$. One must bear in mind, however, that the coherent beam must be sufficienctly strong to ensure the validity of the appoximation.

\begin{figure}
\centering
\includegraphics[width=6cm]{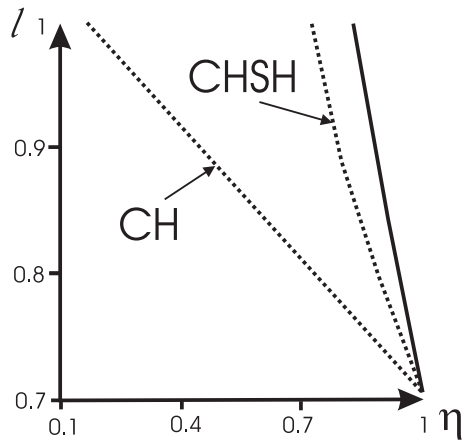}
\caption{Relation between $l_{CRIT}$ and the critical transmission/detection
efficiency $\eta_{CRIT}$ for two-photon (solid line) and
single-photon (dotted line) experiments for the CH and CHSH inequality. Only above the curves,
respectively, the violation of (\ref{ch}) and (\ref{chsh1}) is possible.}
\end{figure}

One should mention here another proposition of this type, posed and experimentally realized by Hessmo {\em et al.} \cite{bjork2}. The most important conceptual
difference between the experiments is that in \cite{bjork2} photons
are not counted, but instead each experimentalist hopes to detect exactly one photon. In the first order of calculus one photon from this pair comes from the coherent beam and the other enters the setup by input $A$. The optimal intensity of the local oscillator
beam is also about one photon per pulse (in front of the detectors), which in the approach from
\cite{Bjork} is not enough to violate the CH inequality. For such a low excitation number our approximation is not valid, and the sum of local probabilities is far less than $1/2$ (see FIG. 3 in \cite{Bjork}).

In conclusion, the threshold for the decoherence parameter looks
similar to the analogous parameter for depolarizing channel acting
on a two-qubit singlet state and producing a Werner state. A
surprising feature of the BJSS scheme is the
critical channel efficiency, see figure 2.
The inequalities are  violated in the right-hand upper corner of the region
of parameters shown in the figure, above the respective curves. For
the non-depolarized case, one has the efficiency threshold
which is much lower than in the standard case of the singlet state
Bell experiment. Non--classicality is carried by one, not two
photons. A loss of the photon has an analogue in a 2-qubit picture of adding a monochromatic product admixture $\ket{00}\bra{00}$ to the entangled state $\ket{\Psi^-}\bra{\Psi^-}$, so that the two states are orthogonal. It is then known by the Peres-Horodeki criterion \cite{Peres} that an arbitrarily small weight of the Bell state in the mixture preserves entanglement. 

Therefore there is a high incentive to perform such an experiment
for sufficiently efficient detectors. However, such an experiment would
additionally require a precise tailoring of the frequency profile of
both the single photon beam and the coherent beam. If there is a
mismatch one cannot expect high visibilities even for non--decohered
single photon beam.

Interestingly, unlike in case of two entangled photons, the CH inequality is not equivalent to the CHSH inequality. As the latter provides a reasonable improvement ($71.8\%$ rather than $82.8\%$), for the former the critical transmission efficiency can be as low as $17.2\%$. However, one needs complicated measurement devices. This is the most challenging aspect for a possible experimental realization. 
Nevertheless, the very high resistance to photon loss makes the proposal of Ref. \cite{Bjork} an attractive scheme for quantum informational applications.

The work is part of EU 6FP programme QAP.
M. \.Zukowski is supported by Wenner-Gren Foundations.   M.
Wie\'sniak is supported by an FNP stipend (within Professorial Subsidy 14/2003 for MZ). The early stage of this work was supported by a UG grant BW 5400-5-0260-4, and a MNiI Grant
1 P03B 04927.

\end{document}